# Unveiling the electrical and thermoelectric properties of highly degenerate indium selenide thin films: Indication of In$_3$Se$_4$ phase

Jaker Hossain*[a], Md. Julkarnain[a], B. K. Mondal[a], M. A. Newaz[a] and K. A. Khan[a]

[a]Solar Energy Laboratory, Department of Electrical and Electronic Engineering, University of Rajshahi, Rajshahi 6205, Bangladesh.

**Abstract**

The effects of annealing and variation of temperature on the electrical and thermoelectric properties of e-beam evaporated InSe thin films has been investigated in details. The XRD study demonstrates that the as-deposited InSe thin films are amorphous while they become polycrystalline with the presence of In$_3$Se$_4$ phase after annealing. The SEM micrographs reveal that the surfaces of as-deposited films are smooth whereas they become non-uniform due to annealing. The heating and cooling cycles of the as-deposited films exhibit that the resistivity of the films shows an irreversible phase-transition and become stable after 3-4 successive heat-treatment operations in air. The electrical conductivity of annealed InSe thin films shows a highly degenerate semiconducting (metallic) behavior. The thermopower of the annealed films indicates that InSe thin film is a highly degenerate n-type semiconductor i.e. metallic. Thickness dependence thermopower obeys the Fuchs-Sondheimer theory. The optical band gap of the annealed films increases as compared to the as-deposited films. These results indicate that InSe thin films encounter a phase-transformation from In$_2$Se$_3$ to a new In$_3$Se$_4$ metallic phase with an optical band gap of ~1.8 eV due to heat-treatment.

## 1. Introduction

Indium selenide (InSe) is a III-VI layered semiconductor made of stacked layers of Se-In-In-Se atoms bonded with van der Waals force [1-2]. The InSe thin films have superior electrical, optical and phase-change properties which have promoted their potential applications in photovoltaic cells [3–6], solid-state batteries [7, 8], phase-change memories [9,10], detectors of ionizing radiation [11], as a suitable candidate for nanoelectronics [12], optoelectronic devices [13, 14] and Field-effect transistors [13]. The InSe-based devices also show excellent flexibility [6, 15] and air stability [16]. Nowadays, many researchers have reported the magnificent performance of InSe as a thermoelectric (TE) power conversion materials [17-19]. The In–Se thin film system can reside in many phases such as InSe, $In_2Se$, $In_2Se_3$, $In_5Se_6$, $In_4Se_3$ and $In_6Se_7$ [20]. Recently, the new $In_3Se_4$ phase of InSe has been reported by some groups having rhombohedral Se-In-Se-In-Se-In-Se layered structure [21-23]. This new 2D-compound with superior properties has widen the opportunity to fabricate new optoelectronic devices in the future [21,23]. More recently, $In_3Se_4$ has been used as a novel anode material for high-performance Li-ion batteries (LIBs) [24].

The InSe thin films have extensively been studied fabricating them by various methods such as thermal evaporation [25-29], flash evaporation [30,31], molecular beam epitaxy [32], chemical vapor deposition [33], Van der Wall epitaxy [34], sol-gel derived techniques [35] and electrodeposition [36]. The maximum of the studies reveal the structural, electrical, electro-optical, electrochemical and spectroscopic properties of the films, phase transformations with temperature and composition, and their potential usefulness in different applications [37-39].

Although the electrical, optical and electrochemical properties of InSe films have been studied and reported by many authors [11, 30-34], no systematic investigation appears to have been performed on electrical and thermoelectric properties by e-beam evaporation technique. Hence, it is necessary to explore the fundamental properties of indium selenide thin films to examine its appropriateness in different device applications.

In this paper, we unveil the effects of annealing and temperature variation on the electrical and thermoelectric properties of the InSe thin films prepared by e-beam evaporation technique and also reveal the presence of the new $In_3Se_4$ phase by interpreting the experimental results to explore the physical properties of the phase as very few works have been done on this material.

## 2. Experimental details

*2.1. Film preparation*

The thin films of InSe was deposited by electron-beam deposition method on glass substrate at a chamber pressure of ~$8 \times 10^{-5}$ Pa. The InSe granular powder was purchased from Metron, USA with a purity of 99.999%. A more detail of the film fabrication procedure can be found in other work [40]. The thickness of the film was measured by an optical Tolansky interference method with a precision of ±5 nm [41]. The film fabrication rate was estimated to 8.30 nms$^{-1}$. The samples were annealed at 475 K for 30 minutes. This temperature is well above the phase-transition temperature for all of the InSe thin films as discussed later.

*2.2. Film Characterization*

The structure of the InSe thin films of different thicknesses was investigated by X-ray diffraction (XRD) technique using the monochromatic CuK$_α$ radiation with Philips "X-pert Pro XRD system". The surface morphology and Energy Dispersive Analysis of X-ray (EDAX) studies of the InSe thin films were carried out by scanning electron microscopy (SEM) with Hitachi S-3400N electron microscope system.

*2.3. Electrical and Thermoelectric studies*

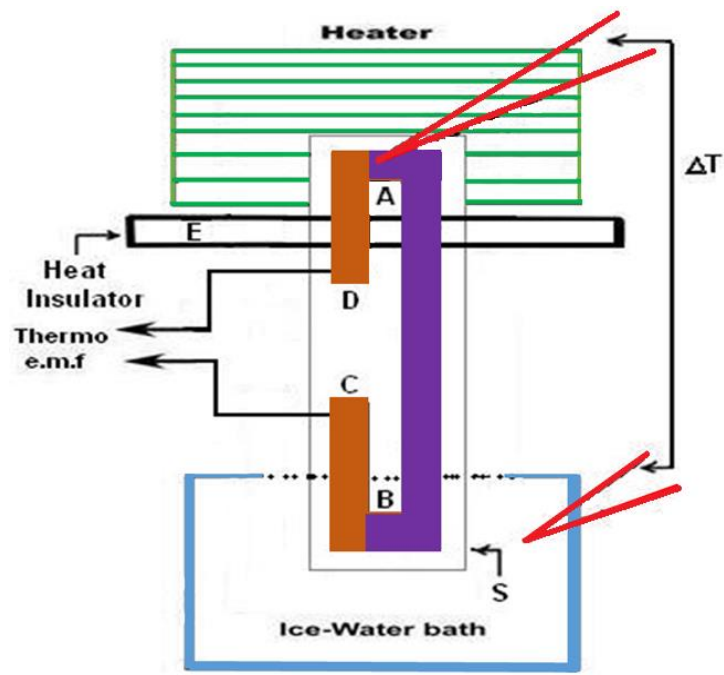

**Fig. 1.** The Schematic diagram of the thermopower measuring system: **AB= InSe** thin film, **AD** and **BC**=Cu film, **S**=substrate, **E**= heat barrier, Δ**T**= the temperature difference between hot and cold junctions.

The electrical study of the InSe thin films was carried out by the Vander Paw method [42]. The Hall effect study of the InSe thin films was also carried out by the Vander Paw method with magnetic field supplied a by electromagnet from Newport Instruments Ltd., England [42]. The thermoelectric power (TEP) measurement of the annealed InSe thin films was carried out by the integral method [43] using a set up as shown in Fig. 1. As shown in the figure, a thick copper film deposited by e-beam technique was used as a reference metal. A flat nichrome strip heater was used to raise the temperature of the junction A. On the other hand, temperature of the junction B was maintained at constant 273 K by keeping it in an ice-water bath. The temperature of the hot junction was measured by a Type K digital thermometer (RS Components, Model: 610-067) connected to the junction. The thermo e.m.f. produced at the system was measured by a digital electrometer (Model: Keithley 614). A heat barrier was used to thermally isolate the hot and cold junctions. An electrical insulation was made to impede any leakage of e.m.f in the dipped portions of the junction B. The temperature of the junction A was increased gradually from room temperature up to 515 K. The measurement error was within ±5%. The system is simple and reliable to develop as a laboratory measurement tool.

## 3. Results and Discussion

*3.1 X-ray diffraction and SEM studies*

(a)

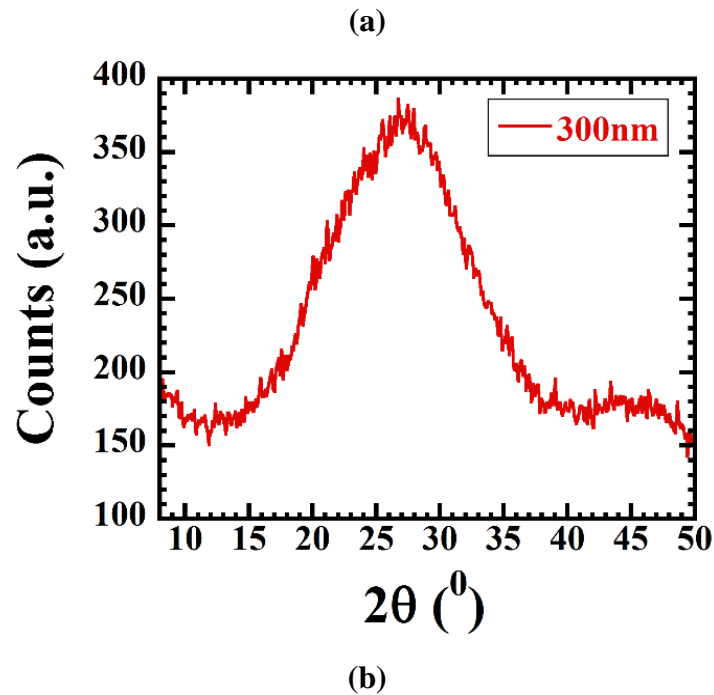

(b)

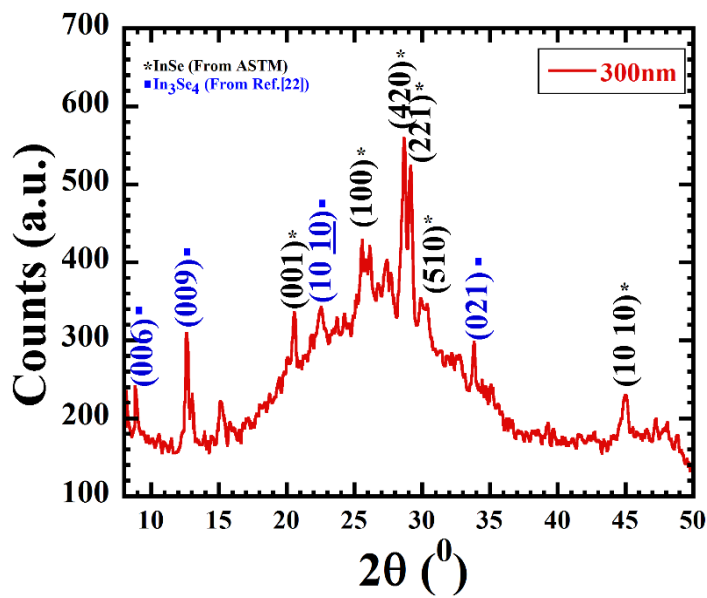

**Fig. 2.** The XRD patterns of (a) as-deposited and (b) annealed InSe thin films of thickness 300 nm.

Fig. 2 (a) and (b) delineate the XRD patterns of the as-deposited and annealed InSe thin films of thickness 300 nm, respectively. The XRD patterns of the 250 nm thick as-deposited and annealed InSe thin films are also shown in Fig. S1 in the supporting information. It is seen from the figures that the as-deposited InSe thin films are amorphous in nature while they are polycrystalline after annealing. The annealed films have the preferential orientations along (001), (100), (101), (420), (221), (510) and (1010) planes as estimated from ASTM cards (PDFs: 00-034-1431, 00-012-0118, 00-012-0118, 00-012-0118 and 00-029-0676), respectively as shown in the figures. The other four peaks that appear at 2θ of 8.782°, 12.651°, 22.466°, and 33.835° can be indexed as (006), (009), (10$\underline{10}$) and (021) planes, respectively comparing the XRD data of the reported work[22]. These results indicate the existence of rhombohedral $In_3Se_4$ phase in the InSe thin films[22]. The XRD data for $In_3Se_4$ are summarized in Table S1 in the supporting information. Therefore, it can be concluded from the XRD results that the as-deposited InSe samples show amorphous behavior whereas the samples become polycrystalline after annealing with the presence of $In_3Se_4$ phase in the films[22, 33, 40, 44-45].

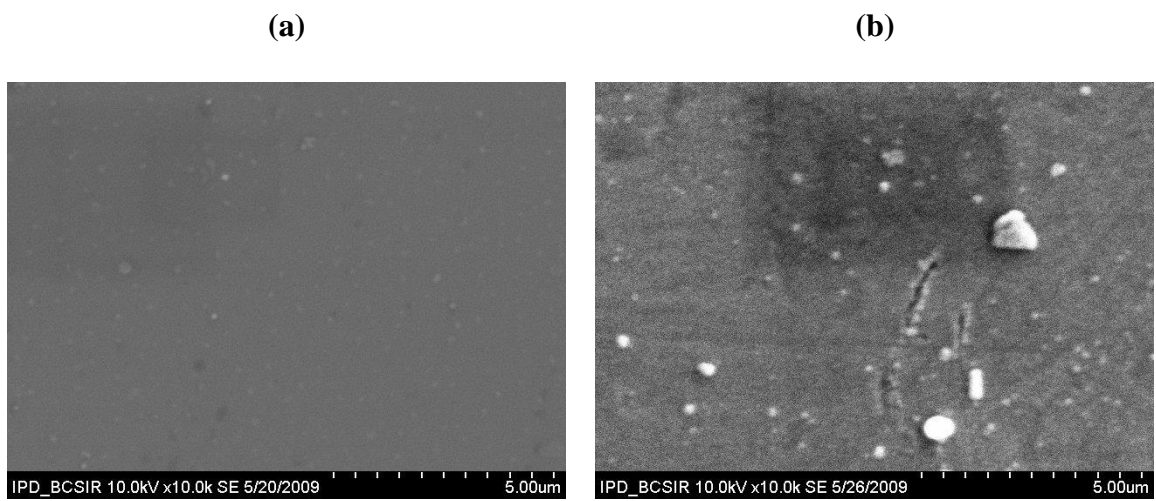

**Fig. 3.** The SEM images of (a) as-deposited and (b) annealed InSe thin films of thickness 300 nm.

The SEM micrographs of the surfaces of 300 nm thick as-deposited and annealed InSe thin films are shown in Fig. 3(a) and (b), respectively. The SEM micrographs of the surfaces of 175 and 250 nm InSe thin films are also shown in Fig. S2 and S3, respectively in the supporting information. It is seen from the figures that the as-deposited films have an almost smooth and uniform surfaces. The samples become nonuniform and contain a few grains on the surfaces due to annealing in air which may arises due to dust or indium-reach dots as selenium deficiency occurs in the annealed films. The data for elemental composition by EDAX of InSe thin films of different thicknesses are shown in Table 1. The EDAX spectra of the corresponding InSe thin films are shown in Fig. S4 and S5 in the supporting information. These results indicate that the as-deposited indium selenide thin films have almost $In_2Se_3$ (In=40at%, Se=60at%) phase whereas the annealed films have nearly $In_3Se_4$ (In=42.86at%, Se=57.14at%) phase of the films.

**Table 1** Elemental compositions by EDAX analysis of indium selenide thin films.

| Film thickness (nm) | Elements | | Error (%) | | Remarks |
|---|---|---|---|---|---|
| | In (at%) | Se (at%) | In | Se | |
| 250 | 39.21 | 60.79 | (±3.3) | (±4.5) | As-deposited |
| | 41.91 | 58.09 | (±3.0) | (±4.6) | Annealed |
| 300 | 40.07 | 59.93 | (±3.2) | (±4.8) | As-deposited |
| | 42.50 | 57.50 | (±4.4) | (±5.1) | Annealed |

## 3.2. Temperature effect on resistivity

Fig. 4 shows the change of resistivity with temperature of a 345 nm thick as-deposited InSe thin film in air. The heat-treatment of the film has been controlled a rate of 5 Kmin$^{-1}$ in every step with a 10 min gap between successive heating cycles.

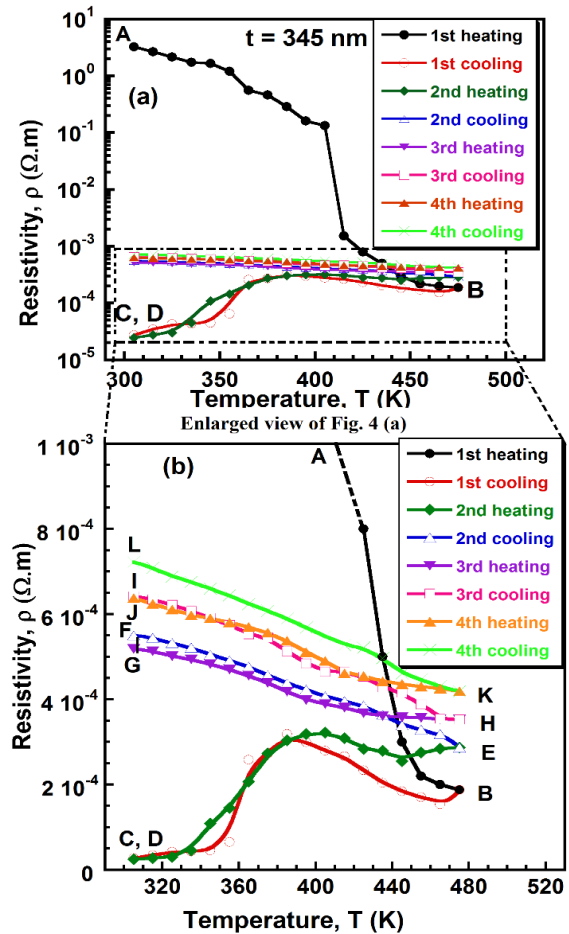

**Fig. 4.** Temperature dependent resistivity of an (a) as-deposited 345 nm thick InSe film and (b) an enlarged view of (a).

As seen in the figure, there is an abrupt fall in resistivity of the InSe thin film in the first heating operation in the observed temperature range of 305-475K. This transition occurs at a temperature of ~415 K and the magnitude of this decrement in resistivity at this particular temperature is about 4.5×10$^3$ order smaller than that of the value at room temperature. It may be noted that after first heating operation, it exhibits a very small variation of resistivity and it overlaps in the investigated temperature range within several successive heat operations. In order to examine the nature of small variation in the overlapping zone, the dot marked zone of Fig. 4(a) is shown in Fig. 4(b) as an enlarged view. The first heating cycle is shown as curve AB in Fig. 4(a). The first cooling operation of the InSe thin film is shown by line BC in Fig. 4(b), displaying the increment in resistivity with the fall of temperature in a different path rather than BA. The second heating operation as indicated by line DE, exhibits that the resistivity softly rises and it follows the same pattern of first cooling operation. The second cooling operation is represented by line EF. It may be noted that resistivity of the InSe thin film increases to about an order of 20 from the beginning of the second heat operation and subsequent chilling to room temperature.

The third subsequent heating operation has also been performed at ambient conditions. A smaller fall of resistivity is observed from point F to G between the end of second chilling operation and beginning of third heating operation. The change of resistivity in the time of third heating and subsequent chilling operations is exhibited by paths GH and HI, respectively. The fourth successive heating and cooling operations are shown by lines JK and KL indicating the decrement or increment of resistivity with temperature. The differences in resistivity among successive heating and chilling operations demonstrate that JL<GI<DF<<<AC. Therefore, investigation on temperature effect on resistivity reveals that the electrical resistivity of InSe thin films become reversible and stable with about $4.5 \times 10^3$ order decrease in resistivity from the as-deposited condition after 3 to 4 consecutive heating and chilling operations.

The first sharp fall in resistivity may happen due to the change in phase of the InSe thin films. The irreversible phase-transition property of the films indicates the changes in the film structure by forming highly conducting $In_3Se_4$ phase during heating-cooling processes and also the improvement in the crystallinity of the film as revealed by XRD study. Similar type of temperature dependent electrical conductivity for InSe thin films has also been reported by other researchers[44,46-47].

*3.3. Electrical properties*

The change in electrical conductivity ($\sigma$) with temperature (T) in the range of 305-475K has been observed for the annealed InSe thin films. Fig. 5 shows the ln$\sigma$ vs. 1/T curves for the four InSe thin films of thickness 100, 150, 200 and 250 nm, respectively. It is seen from the figure that the conductivity increases with temperature which indicates the semiconducting behavior of the films. The ln$\sigma$ vs. 1/T variation of these samples suggests a thermally activated conduction mechanism. The activation energies of these samples have been calculated from the slope of ln$\sigma$ vs. 1/T plots based on the following equation[20]

$$\sigma = \sigma_0 \exp\left(\frac{-\Delta E_0}{K_B T}\right) \quad (1)$$

where $\Delta E_o$ is the activation energy, $\sigma_o$ is pre-exponential factor and $k_B$=Boltzmann constant, respectively. The activation energy of four annealed InSe samples calculated from these plots is shown in Table 2. It is seen from the table that the activation energy is in the range of 36-43 meV which is of the order of $k_B T$, suggesting highly degenerate semiconducting or metallic behavior.

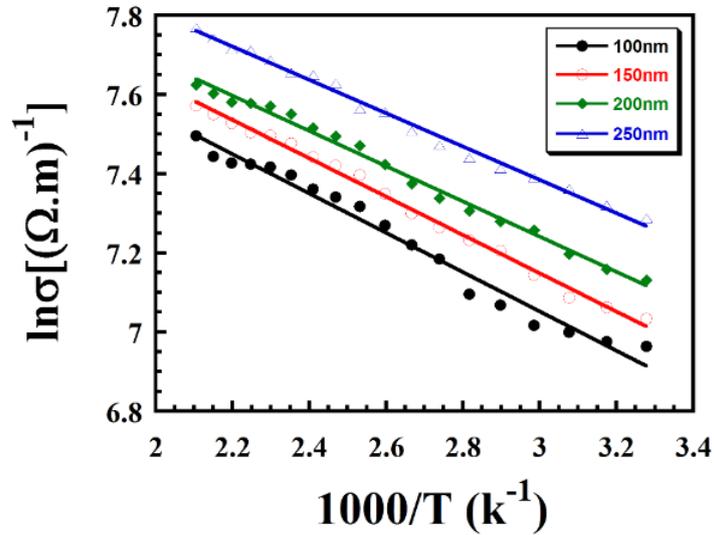

**Fig. 5.** Variation of lnσ vs. inverse temperature for annealed InSe thin films.

**Table 2** Activation energy, $\Delta E_0$ and pre-exponential factor, $\sigma_o$ evaluated for annealed InSe films of different thicknesses.

| Film thickness (nm) | Activation energy, $\Delta E_0$ (meV) | $\sigma_0$ $(\Omega.m)^{-1}$ |
|---|---|---|
| 100 | 43 | 5115.86 |
| 150 | 42 | 5428.94 |
| 200 | 38 | 5294.903 |
| 250 | 36 | 5696.46 |

*3.4. Thermoelectric properties*

The variation of thero e.m.f. and TEP of the annealed InSe thin films with temperature are shown in Fig. 6(a) and (b), respectively. It seen from the Fig. 6(b) that the TEP or Seebeck coefficient (S) is negative which indicates that these thin films are n-type semiconductors. This observation has also been verified by Hall effect study on the thin films. The Hall constant of annealed InSe samples varied in the range of ≈-(1.0-1.6) ×10$^{-3}$ cm$^3$C$^{-1}$ which gives the carrier concentration in the range of range ≈(4-6.5) ×10$^{21}$ cm$^{-3}$. The carrier mobility is of the order of 10$^{-2}$ cm$^2$ V$^{-1}$ s$^{-1}$ for the films. It is also observed from the figure that TEP of the annealed films increases with temperature which reveals a typical degenerate semiconducting behavior of the InSe thin films. The films exhibit a low value of S smaller than -k$_B$/e i.e. -86 µVK$^{-1}$ at the entire temperature range. This situation indicates the highly metallic behavior with the current conducted by electrons with energies within a few multiple of k$_B$T of Fermi level, E$_F$[48]. This is also well consistent with the general concept that the thermopower is smaller when the conductivity of the materials is higher[49].

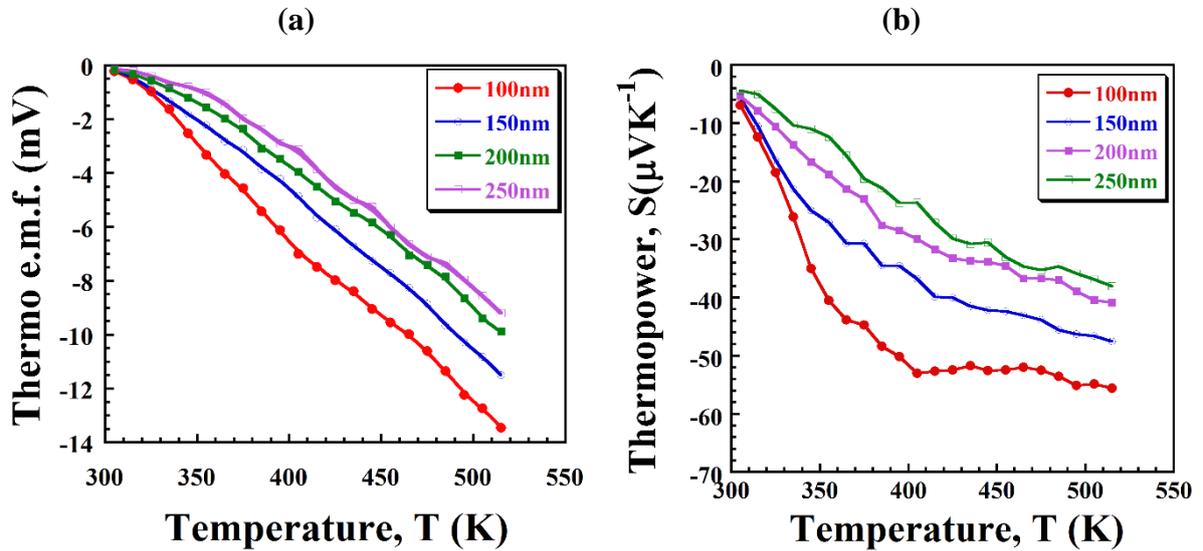

**Fig. 6.** Variation of (a) thermo e.m.f. and (b) thermoelectric power with temperature of annealed InSe thin films.

Seebeck coefficient in metals and degenerate semiconductors is correlated with Fermi energy as[50]

$$S = \frac{8\pi^2 k_B^2 T^2}{3eE_F}\left(\frac{3}{2} + r\right) \quad (2)$$

where S represents the Seebeck coefficient, k$_B$ denotes the Boltzmann constant, T is the absolute temperature, e is the electronic charge, E$_F$ is the Fermi energy and r is the scattering parameter. From the TEP data, r has been calculated to be $\frac{3}{2}$ indicating ionized impurity scattering is dominant in the films[51]. Using the estimated value of S at 305K, the above-mentioned equation provides Fermi energy, $E_F \approx 4.4$ eV.

The effective mass of electron, $m_e^*$ at room temperature has been calculated from the equation [50]

$$S = \frac{8\pi^2 k_B^2 T^2}{3eh^2} m_e^* \left(\frac{\pi}{3n}\right)^{2/3} \qquad (3)$$

where $m_e^*$ indicates the effective mass of electron and n denotes the electron concentration, h is the Planck's constant, respectively. The value of n for the annealed InSe thin films has been measured to be $\approx 4.05 \times 10^{21}$ cm$^{-3}$ by Hall study at room temperature. Putting the estimated value of S and n at T=305K, the equation (3) yields, $m_e^* \approx 0.95 m_0$ where $m_0$ denotes rest mass of electron.

### 3.4.1. Size effect

According to Fuchs-Sondheimer classical size-effect theory, the thermoelectric power S of a polycrystalline thin film of thickness t is given by[52]

$$S = \frac{\pi^2 k_B^2 T}{3eE_F}\left[V + U\frac{\sigma_g}{\sigma_B} - \frac{3}{8}(1-p)\frac{\lambda_B}{t}U\left[\frac{\sigma_g}{\sigma_B}\right]^2\right] \qquad (4)$$

where $V = \left(\frac{dlnA}{dlnE}\right)_{E=E_F}$ is the rate of variation of the Fermi-surface area estimated at the Fermi energy, $E_F$, $U = \left(\frac{dln\lambda_B}{dlnE}\right)_{E=E_F}$ is the rate of change of mean free path with the energy estimated at the $E_F$, e is the charge of the charge carrier, $k_B$ is the Boltzmann constant, and T is the temperature in K, p is the specularity parameter, $\sigma_g$ and $\sigma_B$ are the grain-boundary conductivity and the bulk conductivity, respectively.

It is seen from the above equation that thermopower, S should be a linear function of reciprocal thickness, if p is assumed to be a constant. Fig. 7 shows the variation of thermopower with reciprocal thickness at different temperatures. It is seen from the figure that S is linear with reciprocal thickness at every temperature. So, it can be concluded that InSe thin films obey the inverse thickness dependence predicted by the Fuchs-Sondheimer size-effect theories[52].

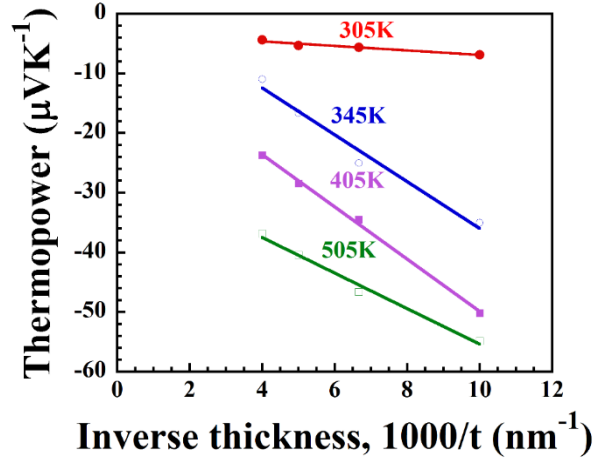

**Fig. 7.** Variation of thermoelectric power with reciprocal thickness of annealed InSe films at temperatures of 305K, 345K 405K and 505K, respectively.

*3.5. Transition from $In_2Se_3$ to $In_3Se_4$ phase*

It has already been reported that the rare-earth chalcogenide goes to phase transformation from $R_2X_3$ to $R_3X_4$, where R and X stand for the rare-earth and chalcogen atoms such as S, Se or Te, respectively[49,53-55]. These materials have the cubic structure. In $R_2X_3$ phase, the two $R^{3+}$ ions provide 6 electrons to the chemical bonding which are taken up by the 3 $X^{2-}$ ions making it insulator as there are no excess electrons available for conduction. On the other hand, in $R_3X_4$ phase nine bonding electrons are provided by three $R^{3+}$ ions, 8 of them are captured by 4 $X^{-2}$ ions giving rise to an excess of one conduction electron per formula unit[49]. The carrier concentration of $R_{3-x}X_4$ phase is given by[53]

$$n = \frac{N_A d}{M}(1 - 3x) \tag{5}$$

where $N_A$ is Avogadro's number, d is the density and M is the molecular weight, respectively. For example, the perfect crystal (x = 0, i.e. no vacancies) of $Ce_3S_4$ has a free electron concentration of $6.25 \times 10^{21}$ cm$^{-3}$ and the x vacancies necessitates (1 - 3x) electrons in the conduction band[53]. Therefore, there is a transition to metallic transport when $R_2X_3$ phase transforms to $R_3X_4$ as the vacancies are filled with rare-earth ions[49,53].

With the same analogy, the metallic behaviour of InSe thin films can be explained. All the investigated as-deposited indium selenide thin films show phase-transformation with transformation temperature roughly depending on film thickness as shown in Fig. S6 in supporting information. As the as-deposited film has $In_2Se_3$ phase (revealed by EDAX data), it is a non-degenerate semiconductor having very low conductivity and optical band gap of

~1.6 eV which is very close to that of β-$In_2Se_3$ [37]. When this phase changes to $In_{3-x}Se_4$ phase (revealed by XRD and EDAX data) with x=0 to 1/3 due to annealing, it become degenerate with high conductivity and optical band gap of ~1.8 eV (Fig. S7 in supporting information)[56]. The carrier concentration of the film is of the order of $10^{21}$ as confirmed by the Hall study which can also be calculated by equation (5) yielding same order. The increase in optical band gap can be calculated from Burstein–Moss theory by the relation[57]

$$\Delta E_g = \frac{h^2}{8m_e^*}\left(\frac{3n}{\pi}\right)^{2/3} \qquad (6)$$

Where h is the plank constant, $m_e^*$ is the effective mass of electron, n is the carrier concentration, respectively. Using the value of $m_e^*$ and n, $\Delta E_g$ is calculated to ~0.38 eV whereas it is 0.2 eV as measured from the optical data. Although, the metallic behaviour of the $In_3Se_4$ phase has already been reported by few authors, the information about thermoelectric and optical band gap of the $In_3Se_4$ material is not confirmed yet[21-23]. For the rare-earth chalcogenides, Zhuze et al. have demonstrated that the Seebeck coefficients and electrical resistivities both increase with increasing temperature and the Fermi levels remain in the conduction band and having a range of 0.4-0.6 eV above the bottom of the conduction band[58]. Our results are also consistent with their observations. As the fermi energy, $E_F$ is ~4.4eV (calculated) and electron affinity of InSe is about 4.55 eV[59], so fermi level will reside within the conduction band as schematically shown in Fig. 8.

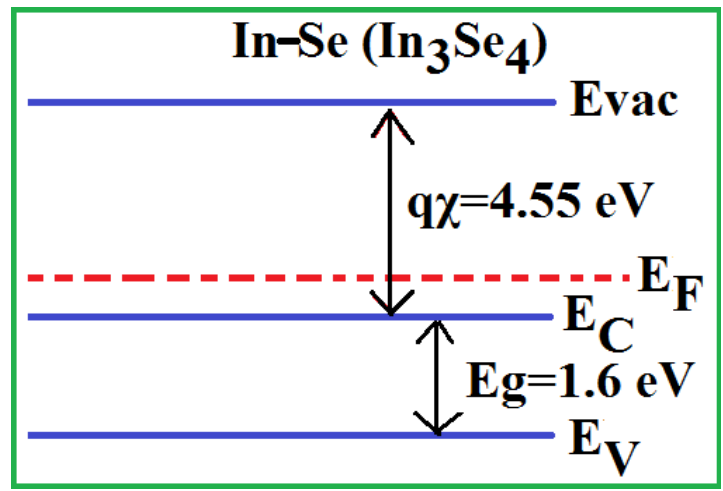

**Fig. 8.** Schematic energy diagram of In-Se thin films showing $In_3Se_4$ phase.

*3.6. Aging effect*

Fig. 9 shows change in electrical conductivity with time for the as-deposited InSe thin films of thicknesses 175 and 250 nm, respectively stored in air at room temperature. It is seen from the figure that the electrical conductivity of the films remains almost stable up to 20 days of aging, and after that, their electrical conductivity increases abruptly. This increase in electrical conductivity may be explained by the mechanism of adsorption or absorption of gases or other impurities to the film. The increase or decrease of conductivity has been found to depend upon the number of molecules adsorbed on the film[60]. These adsorbed molecules may decrease the conductivity of the films if the electron affinity of adsorbed atoms is more than that of the thin films. On the other hand, the conductivity of the films may increase if the films have more electron affinity than that of the adsorbed atoms. The stability and increase of the electrical conductivity of author's sample may be varied due to the latter case.

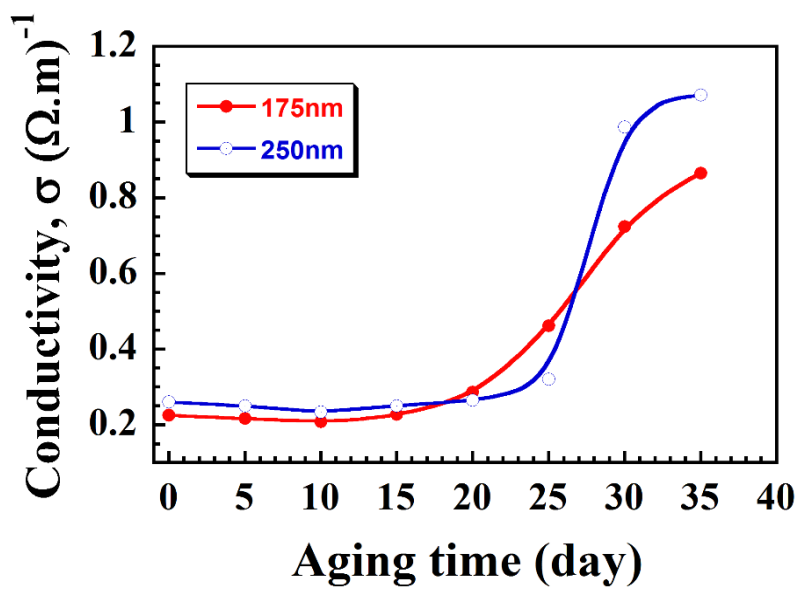

**Fig. 9.** Variation of electrical conductivity with time of as-deposited InSe thin films of thicknesses 175 and 250 nm, respectively.

## 4. Conclusions

The indium selenide thin films have been prepared on glass substrate by e-beam technique. The XRD study unveils that the as-deposited InSe thin films become polycrystalline after annealing with the appearance of $In_3Se_4$ Phase. The EDAX study has been carried out to determine the elemental composition of the films. The temperature variation has remarkable effects on the electrical and thermoelectric properties of InSe thin films. The as-deposited films show an irreversible phase transformation and become stable after 3-4 consecutive operations of heat-treatment in air. The annealed InSe samples shows a degenerate semiconducting behavior. The thermopower study of the annealed films also reveals the highly n-type metallicity of the InSe thin films. Thickness dependence of thermopower follows the Fuchs-Sondheimer size effect theory. The effective mass and fermi energy of the annealed films are found to be $0.95m_0$ and $4.4eV$, respectively. These findings reveal that due to annealing indium selenide thin films goes to a phase change from $In_2Se_3$ to a new $In_3Se_4$ metallic phase with an optical band gap of ~1.8 eV.

**Acknowledgements**

The authors highly appreciate Mrs. Nazia Sultana, and Dr. Samia Tabassum, BCSIR, Dhaka, Bangladesh for their help during the SEM and EDAX studies. The authors also appreciate Dr. Dilip Kumar Saha, AEC, Dhaka, for his support during XRD study of InSe samples. The authors are also indebted to Dr. Mirza H. K. Rubel, Associate Professor, Department of Materials Science and Engineering, University of Rajshahi, Bangladesh for his assistance during the XRD data analysis.

∗**Corresponding author:**

*E-mail address:* jak_apee@ru.ac.bd

- **Associated Content**

**Supporting information**

XRD data, SEM images, EDAX spectra, Phase-transformation temperature as function of film thickness, The plot of $(\alpha h\nu)^2$ vs. $h\nu$ for InSe thin film.

**Declarations of interest:** The authors declare no conflict of interests.

# Supporting Information

**Unveiling the electrical and thermoelectric properties of highly degenerate indium selenide thin films: Indication of In$_3$Se$_4$ phase**

Jaker Hossain[*a], Md. Julkarnain[a], B. K. Mondal[a], M. A. Newaz[a] and K. A. Khan[a]

[a]*Solar Energy Laboratory, Department of Electrical and Electronic Engineering, University of Rajshahi, Rajshahi 6205, Bangladesh.*

*Corresponding author:

 *E-mail address*: jak_apee@ru.ac.bd

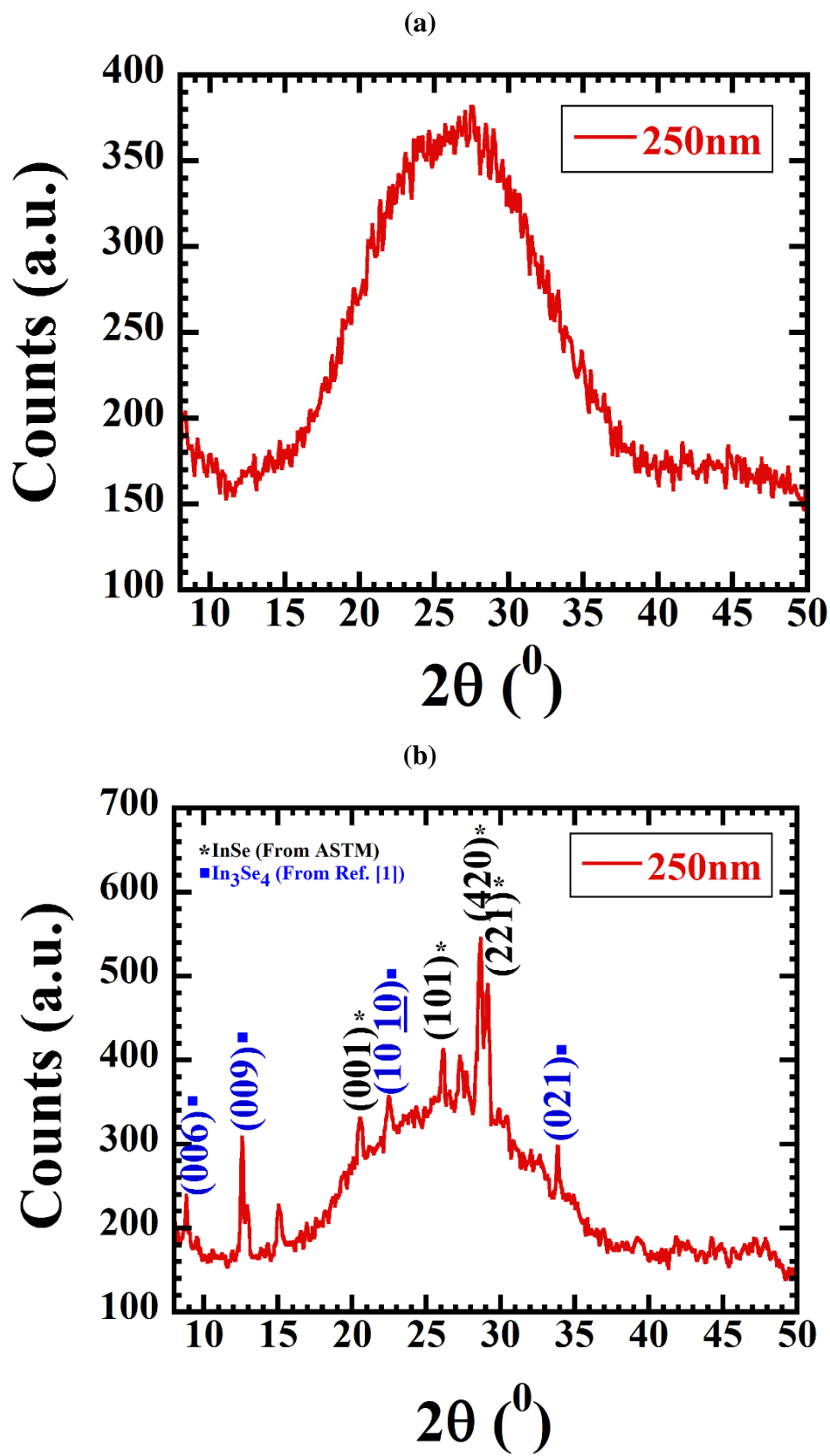

**Fig. S1.** XRD spectra for (a) as-deposited and (b) annealed InSe thin films of thickness 250 nm.

**Table S1** The XRD data for the In$_3$Se$_4$ phase of the annealed indium selenide thin films.

| Thick-ness, t (nm) | Serial No. | 2θ (Deg) | (Observed) d$_{hkl}$ (Å) | Reported 2θ (deg) (Approximate) | Planes (hkl) assigned from reference [1] |
|---|---|---|---|---|---|
| 250 | 1 | 8.8173 | 10.02087 | 8.6 | 006 |
|  | 2 | 12.6246 | 7.01185 | 12.8 | 009 |
|  | 3 | 22.4660 | 3.95760 | 22.2 | 10$\overline{1}$0 |
|  | 4 | 33.8350 | 2.647099 | 33.9 | 021 |
| 300 | 5 | 8.7825 | 10.0605 | 8.6 | 006 |
|  | 6 | 12.6507 | 6.99743 | 12.8 | 009 |
|  | 7 | 22.4704 | 3.95684 | 22.2 | 10$\overline{1}$0 |
|  | 8 | 33.8474 | 2.64619 | 33.9 | 021 |

(a) (b)

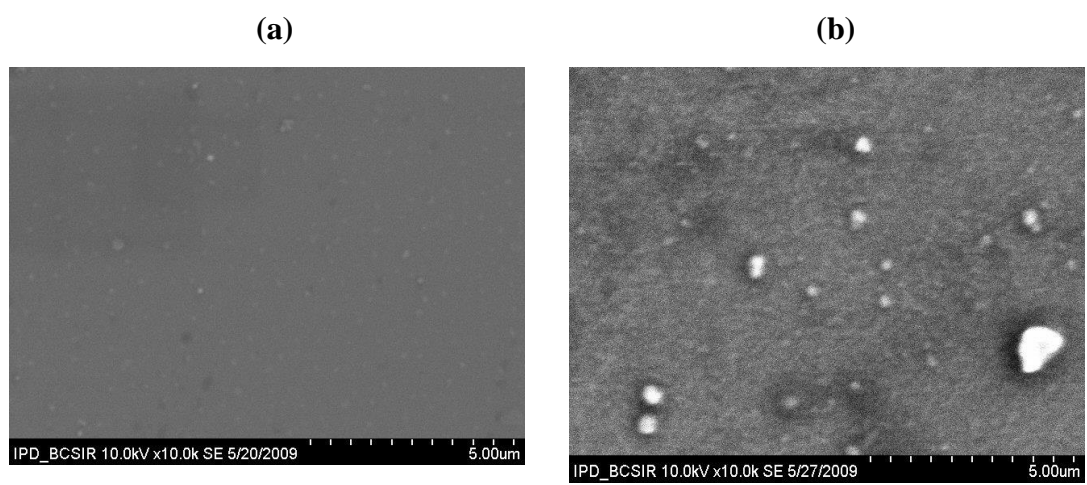

**Fig. S2.** The SEM images of (a) as-deposited and (b) annealed InSe thin films of thickness 175 nm.

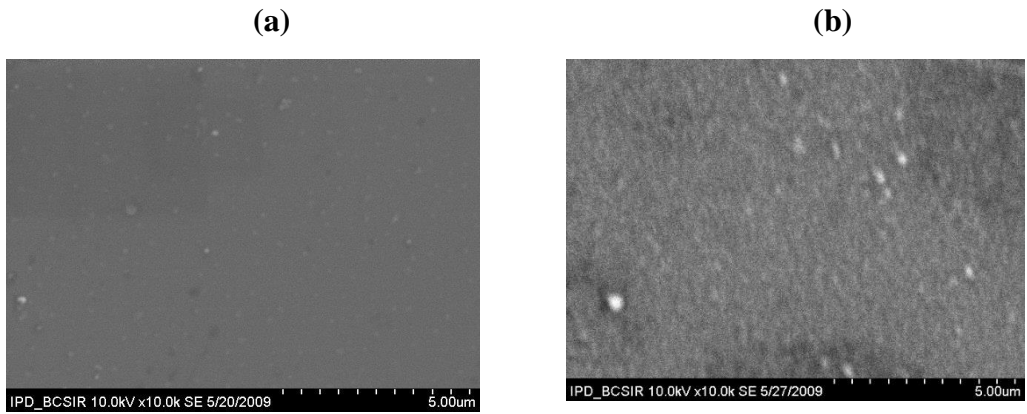

**Fig. S3**. The SEM images of (a) as-deposited and (b) annealed InSe thin films of thickness 250 nm.

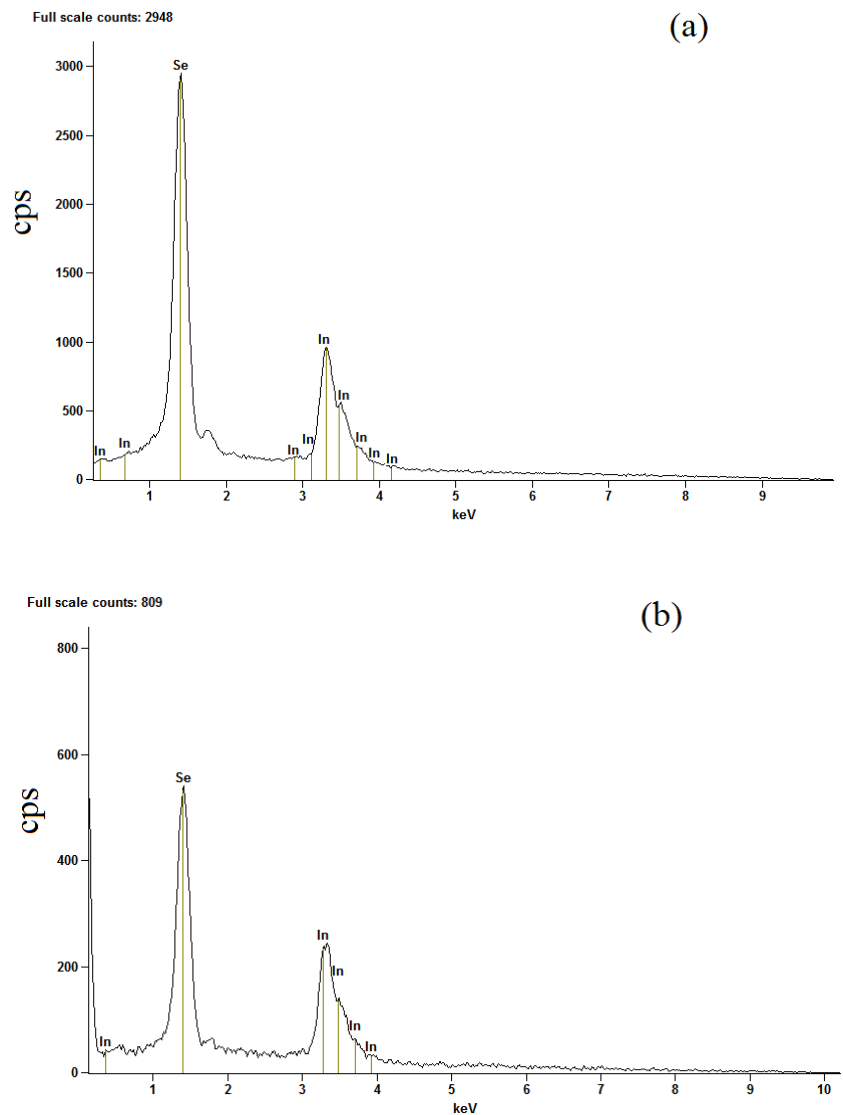

**Fig S4.** EDAX spectra of (a) as-deposited and (b) annealed 250 nm thick indium selenide thin films.

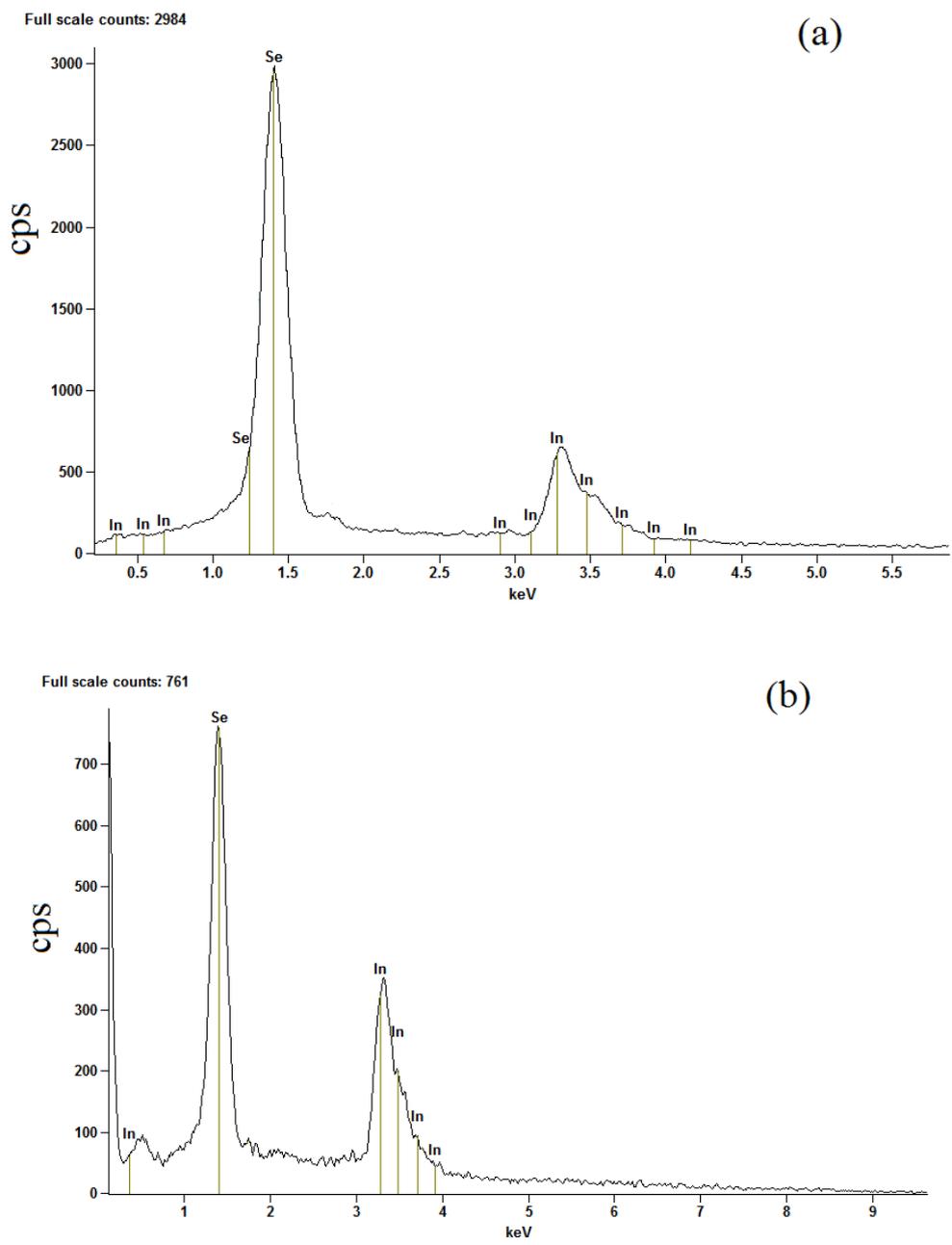

**Fig S5.** EDAX spectra of (a) as-deposited and (b) annealed 300 nm thick indium selenide thin films.

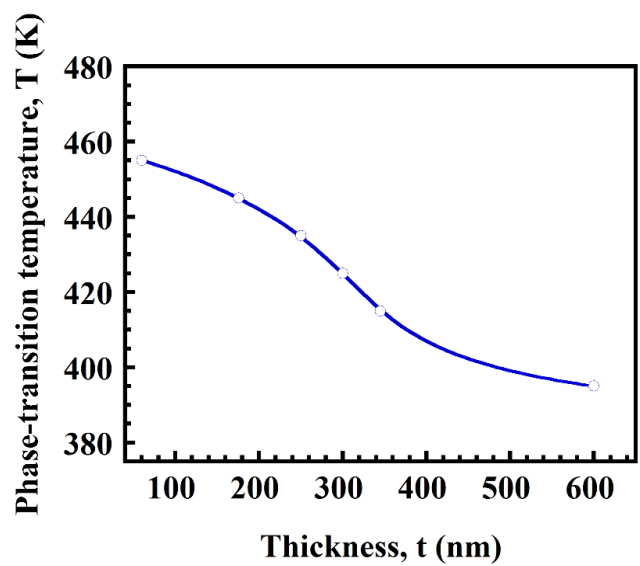

**Fig. S6.** Phase-transformation temperature as function of film thickness of as-deposited InSe thin films.

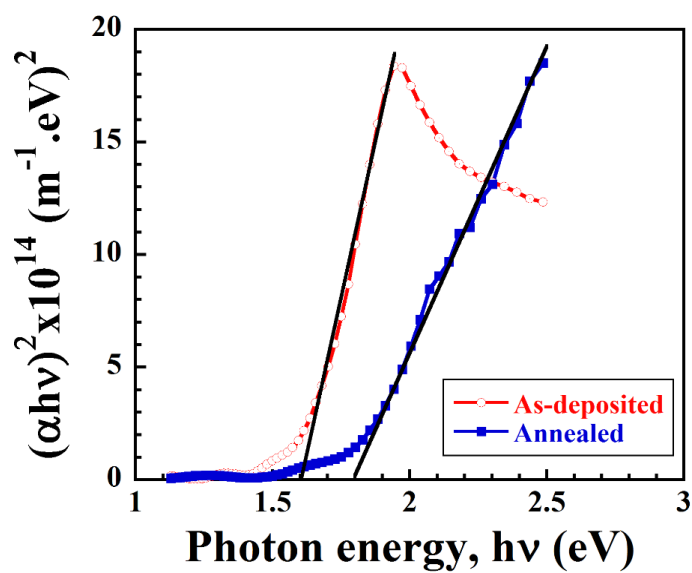

**Fig. S7.** The plot of $(\alpha h\nu)^2$ vs. $h\nu$ for as-deposited and annealed InSe thin film of thickness of 250 nm.

# Reference

[1] G. Han, Z.-G. Chen, C. Sun, L. Yang, L. Cheng, Z. Li, W. Lu, Z.M. Gibbs, G.J. Snyder, K. Jack, J. Drennan J. Zou, CrystEngComm 16 (2014) 393-398.